\documentclass[prl,aps,twocolumn,showpacs,nobibnotes,epsf]{revtex4}

\usepackage{graphicx}
\usepackage{dcolumn}
\usepackage{bm}
\usepackage{SIunits}
\usepackage{verbatim}
\usepackage{placeins}
\usepackage{multirow}
\begin{document}
\title{Evolution of Structural and Physical Properties of Sr$_3$(Ru$_{1-x}$Mn$_x$)$_2$O$_7$ with Mn Concentration}

\author{Biao Hu$^1$, Gregory T. McCandless$^2$, V. O. Garlea$^3$, S. Stadler$^1$, Yimin Xiong$^1$, Julia Y. Chan$^2$, E. W. Plummer$^1$}
\author{R. Jin$^{1,}$}

\email{rjin@lsu.edu}

\affiliation{$^1$Department of Physics $\&$ Astronomy, Louisiana
State University, Baton Rouge, Louisiana 70803, USA}
\affiliation{$^2$Department of Chemistry, Louisiana
State University, Baton Rouge, Louisiana 70803, USA}
\affiliation{$^3$Neutron Scattering Sciences Division, Oak Ridge National Laboratory, Oak Ridge, Tennessee 37831, USA}

\begin{abstract}
Layered ruthenates are prototype materials with strong structure-property correlations. We report the structural and physical properties of double-layered perovskite Sr$_3$(Ru$_{1-x}$Mn$_x$)$_2$O$_7$ single crystals with 0$\leq$\emph{x}$\leq$0.7. Single crystal x-ray diffraction refinements reveal that Mn doping on the Ru site leads to the shrinkage of unit-cell volume and disappearance of (Ru/Mn)O$_6$ octahedron rotation when \emph{x}$>$0.16, while the crystal structure remains tetragonal. Correspondingly, the electronic and magnetic properties change with \emph{x}. The electrical resistivity reveals metallic character ($d\rho/dT$$>$0) at high temperatures but insulating behavior ($d\rho/dT$$<$0) below a characteristic temperature $T_{\rm MIT}$. Interestingly, $T_{\rm MIT}$ is different from $T_{\rm M}$, at which magnetic susceptibility reaches maximum. $T_{\rm MIT}$ monotonically increases with increasing \emph{x} while $T_{\rm M}$ shows non-monotonic dependence with \emph{x}. The difference between $T_{\rm MIT}$ and $T_{\rm M}$ ($T_{\rm MIT}$$>$$T_{\rm M}$) becomes larger when \emph{x}$>$0.16. The constructed phase diagram consists of five distinct regions, demonstrating that the physical properties of such a system can easily be tuned by chemical doping.
\end{abstract}

\vskip          15          pt 

\maketitle

Transition metal oxides (TMOs) have attracted extensive attention due to the strong correlations between charge, lattice, orbital, and spin degrees of freedom. The Ruddleson-Popper (RP) Sr$_{n+1}$Ru$_n$O$_{3n+1}$ (\emph{n}=integer) series are prototype strongly correlated systems, since both theoretical and experimental investigations indicate intimate relationships between structural, electronic and magnetic properties \cite{Maeno, Ikeda, Fobes, Cao, Fang}. A small change in structure often results in different ground states, as seen in single-layered (\emph{n}=1) Ca$_{2-x}$Sr$_{x}$RuO$_4$ \cite{Nakatsuji, Friedt}. Different from the rest of the RP series, Sr$_3$Ru$_2$O$_7$ (\emph{n}=2) shows unique physical properties. Although the electrical resistivity varies smoothly with temperature without any anomaly, the magnetic susceptibility of Sr$_3$Ru$_2$O$_7$ reveals a characteristic peak around 16 K \cite{Ikeda}. Neutron scattering measurements confirm that the susceptibility peak corresponds to a short-range antiferromagnetic (AFM) correlation \cite{Capogna, Stone}. Under the application of hydrostatic pressure, the ground state of Sr$_3$Ru$_2$O$_7$ reveals ferromagnetic (FM) instability \cite{Ikeda}. On the other hand, the application of magnetic field leads to a metamagnetic transition at low temperatures \cite{Perry}. These phenomena strongly suggest that both AFM and FM interactions exist in Sr$_3$Ru$_2$O$_7$.

It was reported that a slight substitution of Ru by Mn drives the ground state from a paramagnetic metal (PM) to an AFM insulator, and a phase diagram of Sr$_3$(Ru$_{1-x}$Mn$_x$)$_2$O$_7$ was mapped out up to \emph{x}=0.2 \cite{Mathieu}. The central question is how Mn doping leads to the change of ground state properties. X-ray absorption spectroscopy (XAS) revealed that the Mn dopant has an oxidation state different from Ru$^{4+}$, while x-ray photoemission spectroscopy (XPS) showed no sign of doping-induced multiple Ru valences up to \emph{x}=0.2 \cite{Hossain, Guo}. We have studied Sr$_3$(Ru$_{1-x}$Mn$_x$)$_2$O$_7$ in the doping range of 0$\leq$\emph{x}$\leq$0.7. According to its electronic and magnetic properties, a phase diagram is constructed which has two phase boundaries: one is a metal-insulator crossover line and the other is the magnetic transition line.

Single crystals of Sr$_3$(Ru$_{1-x}$Mn$_x$)$_2$O$_7$ (0$\leq$\emph{x}$\leq$0.7) were grown by the floating-zone technique in an image furnace (model: Canon SC1-MDH20020). All selected crystals for physical property measurements in this Letter were characterized by powder and single crystal x-ray diffraction (XRD). The crystal structure and Mn concentration (\emph{x}) were determined by single crystal XRD refinement. Magnetic susceptibility measurements were carried out in a superconducting quantum interference device (SQUID) magnetometer. Measurements of the resistivity and specific heat were performed in a Quantum Design Physical Properties Measurement System (PPMS).

For all Sr$_3$(Ru$_{1-x}$Mn$_x$)$_2$O$_7$ samples, single crystal XRD data show that their structure can be described by the space group \emph{I}4/\emph{mmm} with the details described previously \cite{Hu}. The left panel of Fig. 1 displays the unit-cell representation of Sr$_3$(Ru$_{1-x}$Mn$_x$)$_2$O$_7$ (top) and the three oxygen sites of the (Ru/Mn)O$_6$ octahedron (bottom). Fig. 1(a)-(d) shows the \emph{x} dependence of lattice parameters \emph{a} and \emph{c}, the volume of unit cell (\emph{V}), and ratio \emph{c/a} at 298 K and 90 K, respectively. Note that, with increasing \emph{x}, the lattice parameter \emph{a} increases for 0$\leq$\emph{x}$\leq$0.2 and decreases for \emph{x}$>$0.2, while the lattice parameter \emph{c} decreases monotonically. This results in a monotonic decrease of \emph{V} and \emph{c/a} with increasing \emph{x}. For comparison, the structural information obtained from polycrystalline Sr$_3$Mn$_2$O$_7$ (\emph{x}=1) is also presented \cite{Mitchell}, which has a higher \emph{c/a} ratio than that for \emph{x}=0.7. Nevertheless, the Ru-O(3) bond length [Fig. 1(g)] remains more or less unchanged, while both the Ru-O(1) [Fig. 1(e)] and Ru-O(2) bond lengths [Fig. 1(f)] decrease with increasing \emph{x}. With this information, the Jahn-Teller distortion ($\Delta_{\rm JT}$) can be calculated via $\Delta_{\rm JT}$=[Ru-O(1)+Ru-O(2)]/[2$\times$Ru-O(3)], which decreases from 1.04 for \emph{x}=0 to 1 for \emph{x}=0.7 (not shown). This suggests that Mn doping makes the (Ru/Mn)O$_6$ octahedron less distorted. Further support can be found from the reduction of rotation angle of the (Ru/Mn)O$_6$ octahedron, as shown in Fig. 1(h). Note that the rotation angle ($\Phi$) of the (Ru/Mn)O$_6$ octahedron decreases with increasing \emph{x} and becomes undetectable for \emph{x}$>$0.16.

\begin{figure}[t]
\includegraphics[width=0.50\textwidth]{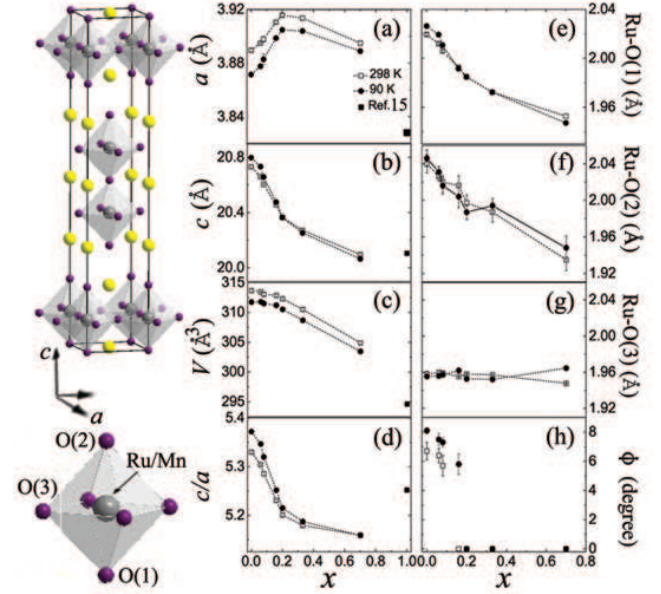}
\caption{(Color online) Unit-cell representation of Sr$_3$(Ru$_{1-x}$Mn$_x$)$_2$O$_7$ in space group \emph{I}4/\emph{mmm} (left top) and the configuration of the (Ru/Mn)O$_6$ octahedron (left bottom), where the Mn atoms partially occupy the Ru site. (a)-(d) are the Mn concentration (\emph{x}) dependence of the lattice parameters \emph{a} and \emph{c}, the unit cell volume (\emph{V}), and the ratio \emph{c/a} at 298 K (empty squares) and 90 K (solid circles), respectively. The solid square in (a)-(d) indicate the values of \emph{a}, \emph{c}, \emph{V} and \emph{c/a} of polycrystalline Sr$_3$Mn$_2$O$_7$ at 300 K obtained from Ref.15. (e)-(g) are the bond length of Ru-O(1) (inner apical), Ru-O(2) (outer apical), and Ru-O(3) (equatorial) as a function of \emph{x} at 298 K (empty squares) and 90 K (solid circles), respectively. (h) is the \emph{x} dependence of the rotation angle $\Phi$ of the (Ru/Mn)O$_6$ octahedron at 298 K and 90 K, respectively. Dashed lines are guides for the eye.}
\end{figure}

The temperature dependence of the in-plane ($\rho_{\rm ab}$) and out-of-plane ($\rho_{\rm c}$) electrical resistivity of Sr$_3$(Ru$_{1-x}$Mn$_x$)$_2$O$_7$ single crystals are shown in Fig. 2(a) and 2(b), respectively. For the undoped compound (\emph{x}=0), both $\rho_{\rm ab}$(T) and $\rho_{\rm c}$(T) are metallic in the measured temperature range. Upon doping, both $\rho_{\rm ab}$ and $\rho_{\rm c}$ are not only enhanced in magnitude but also change sign in slope at a characteristic temperature $T_{\rm MIT}$ from positive (metallic) at high temperatures to negative (insulating) at low temperatures. This result is consistent with the previous report that a metal-insulator transition (MIT) occurs when introducing the Mn dopant into Sr$_3$Ru$_2$O$_7$ \cite{Mathieu}. With increasing \emph{x}, $T_{\rm MIT}$ is quickly pushed to higher temperature and becomes less pronounced.

However, the magnetic properties of Sr$_3$(Ru$_{1-x}$Mn$_x$)$_2$O$_7$ reveal a different trend. Fig. 2(c) and 2(d) show the temperature dependence of the in-plane ($\chi_{\rm ab}$) and out-of-plane ($\chi_{\rm c}$) magnetic susceptibility under zero-field-cooling (ZFC) condition for Sr$_3$(Ru$_{1-x}$Mn$_x$)$_2$O$_7$, respectively ($\chi_{\rm ab}$ and $\chi_{\rm c}$ measured under field-cooling condition are very similar). For 0$\leq$\emph{x}$\leq$0.7, both $\chi_{\rm ab}$ and $\chi_{\rm c}$ always display a characteristic peak at $T_{\rm M}$. For Sr$_3$Ru$_2$O$_7$ (\emph{x}=0), $T_{\rm M}$ is about 16 K, in agreement with previous results \cite{Ikeda}. With increasing \emph{x}, $T_{\rm M}$ initially increases then decreases, with a maximum near \emph{x}$\sim$0.16.

In order to understand why $T_{\rm M}$ varies with \emph{x} nonmonotically, we analyze $\chi_{\rm ab}$ and $\chi_{\rm c}$ at high temperatures. Both $\chi_{\rm ab}$(T) and $\chi_{\rm c}$(T) can be fitted with a formula $\chi$(T)=$\chi_{\rm 0}$+$\chi_{\rm CW}$(T) between 175 K and 390 K. Here $\chi_{\rm 0}$ is the temperature independent term and $\chi_{\rm CW}$(T)=\emph{C}/(T-$\Theta_{\rm CW}$) is the Curie-Weiss term with Curie constant $\emph{C}=N_{\rm A}\emph{p}_{\rm eff}^{2}\mu_{\rm B}^{2}/(3k_{\rm B})$ and Curie-Weiss temperature $\Theta_{\rm CW}$ ($N_{\rm A}$ is Avogadro number, $p_{\rm eff}$ is the effective Bohr magneton number, $\mu_{\rm B}$ is the Bohr magneton, and $k_{\rm B}$ is the Boltzmann constant). $\Theta_{\rm CW}$ and $p_{\rm eff}$ obtained from the fitting for 0$\leq$\emph{x}$\leq$0.7 are plotted in Fig. 2(e) and Fig. 2(f), respectively. Note that both $\Theta_{\rm CW}^{ab}$ and $\Theta_{\rm CW}^{c}$ are negative with similar magnitude and increase with increasing \emph{x} for 0$\leq$\emph{x}$\leq$0.16. For \emph{x}$>$0.2, $\Theta_{\rm CW}^{ab}$ is positive but $\Theta_{\rm CW}^{c}$ is negative. The sign change of $\Theta_{\rm CW}^{ab}$ is likely caused by the change from AFM to FM interaction in the \emph{ab} plane, while the dominant magnetic interaction in \emph{c} direction remains AFM ($\Theta_{\rm CW}$$<$0). Indeed, the in-plane magnetization ($\emph{M}_{\rm ab}$) vs field (H) plot shows FM character when \emph{x}$>$0.16 (see Fig. 2(g)).

Although $\rho_{\rm c}$$\gg$$\rho_{\rm ab}$ with $\rho_{\rm c}$/$\rho_{\rm ab}$$\sim$30 for \emph{x}=0 and $\sim$10 for \emph{x}$>$0.16 at room temperature, the magnetic anisotropy is much smaller with $\chi_{\rm c}$/$\chi_{\rm ab}$$\sim$1 for \emph{x}$>$0.16 (see Fig. 2(h)). The above fitting also showed that $p_{\rm eff}^{ab}$$\sim$$p_{\rm eff}^{c}$. Interestingly, both $p_{\rm eff}^{ab}$ and $p_{\rm eff}^{c}$ increase with \emph{x} and tend to saturate for \emph{x}$>$0.16. For \emph{x}=0, $p_{\rm eff}$$\sim$2.8, corresponding to \emph{S}=1, according to $p_{\rm eff}$=\emph{g}$\sqrt{\emph{S}(\emph{S}+1)}$ with \emph{g}=2 for transition metals. For \emph{x}$>$0.16, $p_{\rm eff}$$\sim$3.7, corresponding to \emph{S}=3/2.

\begin{figure}[t]
\includegraphics[width=0.50\textwidth]{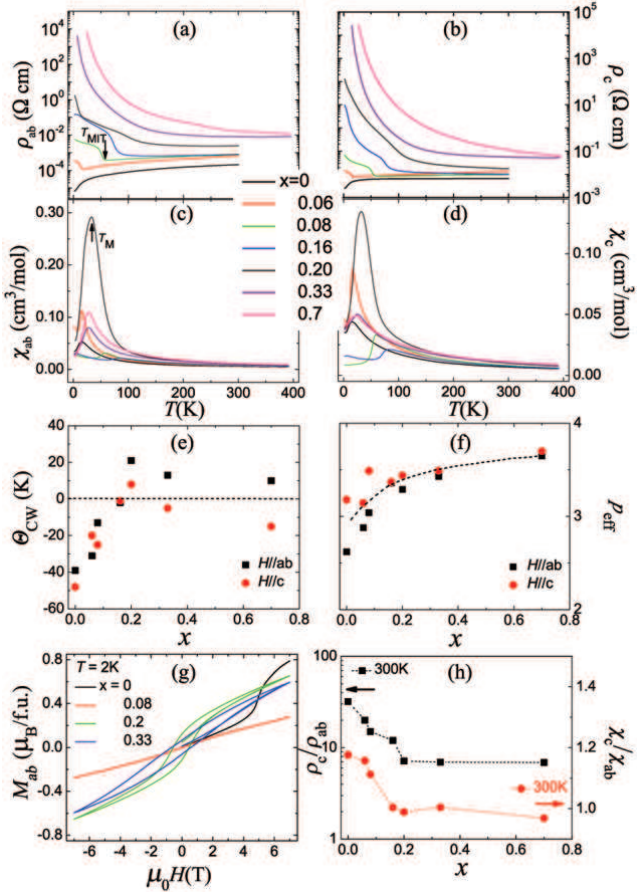}
\caption{(Color online) (a) and (b) are the temperature dependence of $\rho_{\rm ab}$(T) and $\rho_{\rm c}$(T) with different \emph{x}, respectively. The arrow in (a) shows an example of the temperature defined as $T_{\rm MIT}$ for \emph{x}=0.08. (c) and (d) are $\chi_{\rm ab}$(T) and $\chi_{\rm c}$(T) as a function of \emph{T} with different \emph{x}, respectively. The arrow in (c) shows an example of the temperature defined as $T_{\rm M}$ for \emph{x}=0.2. (e) and (f) are the derived $\Theta_{\rm CW}$ and $p_{\rm eff}$ from Curie-Weiss law fitting as a function of \emph{x}, respectively. Squares (circles) denote fitting values under H$\parallel$ab (H$\parallel$c). (g) shows magnetization \emph{M\rm(H)} hysteresis loops at 2 K for H$\parallel$ab for \emph{x}=0 (black), 0.08 (red), 0.2 (blue) and 0.33 (green). (h) shows the \emph{x} dependence of $\rho_{\rm c}$/$\rho_{\rm ab}$ (left axis) and $\chi_{\rm c}$/$\chi_{\rm ab}$ (right axis) at 300 K. Dashed lines are guides for the eye.}
\end{figure}

The temperature dependence of the resistivity and magnetic susceptibility reveals two characteristic temperatures ($T_{\rm MIT}$ and $T_{\rm M}$) in Sr$_3$(Ru$_{1-x}$Mn$_x$)$_2$O$_7$. The question is whether they correspond to true phase transitions. The specific heat data shown in Fig. 3(a) allow us to determine the nature of $T_{\rm MIT}$ and $T_{\rm M}$. In Fig. 3(a), we plot the specific heat as $C_{\rm p}/\emph{T}$ vs \emph{T}, and shift the data for each doping level for clarity. For \emph{x}=0, $C_{\rm p}$ varies with \emph{x} smoothly without any anomaly at $T_{\rm M}$$\sim$16 K (see Fig. 2c). This indicates that there is no true phase transition in the undoped compound, consistent with neutron scattering measurements \cite{Capogna}. For \emph{x}=0.06, there is a clear specific heat anomaly at $T_{\rm M}$, indicating a true second order phase transition. Since $T_{\rm M}$$\sim$$T_{\rm MIT}$ for \emph{x}=0.06, it is unclear whether the phase transition originates from magnetic ordering and/or a metal-insulator transition. Specific heat data for higher doping levels can clarify this. Note that, for \emph{x}=0.16, the specific heat anomaly presents at $T_{\rm M}$$\sim$80 K but not at $T_{\rm MIT}$$\sim$140 K. This implies that $T_{\rm M}$ in the region of 0.06$\leq$\emph{x}$\leq$0.16 corresponds to a true phase transition, while $T_{\rm MIT}$ represents a crossover temperature from metallic behavior at high temperatures to insulating character at low temperatures. Recent neutron scattering experiment confirms a long-range AFM ordering below $T_{\rm M}$ for \emph{x}=0.16 \cite{Mesa}. Theoretically, the entropy removal upon magnetic ordering is expected to be $S_{\rm M}$=\emph{R}ln(2$\emph{S}$+1)=1.09\emph{R} for \emph{S}=1 and 1.39\emph{R} for \emph{S}=3/2 (\emph{R}=8.314 J/mol K). We may estimate the actual entropy removal at $T_{\rm M}$ by subtracting the background by fitting the experimental data outside of the transition region using a polynomial (dashed line in Fig. 3(a)). By integrating $\Delta$$C_{\rm p}$/T in the transition region, we obtain $\Delta$$\emph{S}_{\rm M}$$\sim$0.077\emph{R} for \emph{x}=0.06, 0.64\emph{R} for \emph{x}=0.08, and 0.77\emph{R} for \emph{x}=0.16. These values are considerably smaller than the expected values, indicating that only a fraction of the spins are ordered. It is also possible that some of entropy has been removed above $T_{\rm M}$. Nevertheless, the specific heat anomaly at $T_{\rm M}$ can no longer be detected when \emph{x}$>$0.16 (see Fig. 3(a)), suggesting that there is no long-range magnetic ordering at high Mn doping levels.

As shown in the inset of Fig. 3(b), the low temperature (2 K) specific heat decreases with increasing \emph{x}, quickly dropping to a very small value as \emph{x}$>$0.16. This is most likely due to the reduction of electronic specific heat, because of the insulating ground state when \emph{x}$\neq$0. The electronic specific heat becomes negligible at high Mn doping concentrations. However, the low temperature (below 10 K) $C_{\rm p}/T$ does not seem to vary linearly with $\emph{T}^{2}$ (Fig. 3(b)). Such a deviation should be attributed to magnetic contributions of the system.

\begin{figure}[t]
\includegraphics[width=0.5\textwidth]{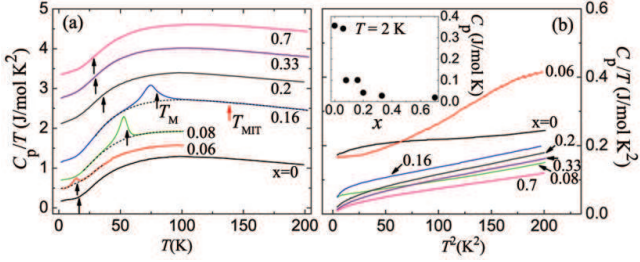}
\caption{(Color online) (a) Temperature dependence of specific heat $C_{\rm p}$ of Sr$_3$(Ru$_{1-x}$Mn$_x$)$_2$O$_7$ at zero field, plotted as $C_{\rm p}(T)/T$ versus \emph{T} and shifted for clarity. The black arrows indicate $T_{\rm M}$ for each concentration. The red arrow indicates $T_{\rm MIT}$ for \emph{x}=0.16. Dashed lines in the \emph{x}=0.06, 0.08 and 0.16 plots represent the polynomial fit to the specific heat background. (b) Low temperature specific heat plotted as $C_{\rm p}/T$ versus $\emph{T}^{2}$. The inset of (b) shows $C_{\rm p}$ at 2 K for each \emph{x}.}
\end{figure}

Based on the above observations, we construct a phase diagram for Sr$_3$(Ru$_{1-x}$Mn$_x$)$_2$O$_7$, covering 0$\leq$\emph{x}$\leq$0.7. Fig. 4 shows the \emph{x}-\emph{T} phase diagram, which consists of two boundary lines: $T_{\rm MIT}$ and $T_{\rm M}$. In terms of physical properties, it can be divided into five regions, as marked in the phase diagram. Region I represents a paramagnetic (PM) metallic (PM-M) phase, which covers temperature range above $T_{\rm MIT}$. Region II is a PM insulating (PM-I) phase, where the system is non-metallic with $d\rho/dT$$<$0 but remains paramagnetic. Region III (0$\leq$\emph{x}$<$0.06) represents metallic phase with AFM correlation (AFMC-M), where the correlation is enhanced upon Mn doping. Region IV is a long-range (LR) AFM insulating (LR-AFM-I) phase, where LR AFM ordering forms below $T_{\rm M}$ and the specific heat anomaly emerges at $T_{\rm M}$. Since there is lack of specific heat anomaly in Region V, this region is an insulating phase with short-range magnetic correlations (SRMC-I).

\begin{figure}[t]
\includegraphics[width=0.5\textwidth]{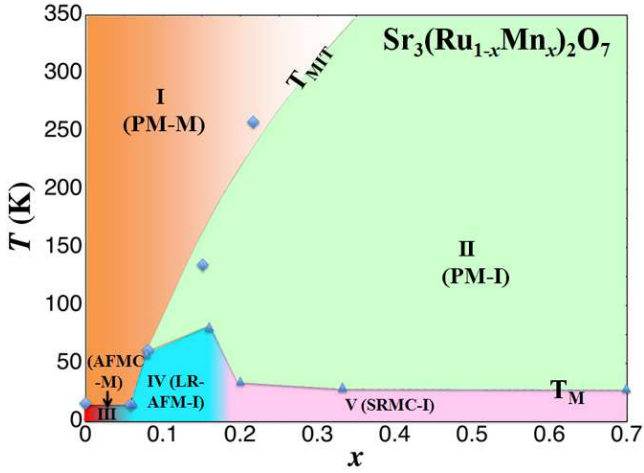}
\caption{(Color online) The \emph{x}-\emph{T} phase diagram of Sr$_3$(Ru$_{1-x}$Mn$_x$)$_2$O$_7$ (0$\leq$\emph{x}$\leq$0.7). Diamonds and triangles represent $T_{\rm MIT}$ and $T_{\rm M}$, respectively. Region I is a paramagnetic metallic (PM-M) phase. Region II is a paramagnetic insulating (PM-I) phase. Region III is a metallic phase with AFM correlation (AFMC-M). Region IV represents a long-range AFM insulating phase (LR-AFM-I). Region V is an insulating phase with short-range magnetic correlation (SRMC-I).}
\end{figure}

In light of all of the structural and physical properties, it becomes clear that the variation of electronic and magnetic properties is intimately connected with the change of structure of Sr$_3$(Ru$_{1-x}$Mn$_x$)$_2$O$_7$. Due to partial replacement of Ru by Mn with smaller ionic radius, the unit cell becomes smaller and (Ru/Mn)O$_6$ becomes less distorted. This is reflected in both rotation angle (see Fig. 1 (h)) and $\Delta_{\rm JT}$ (not shown). This weakens the FM interaction, according to the theoretical calculations for the single layered ruthenate Ca$_{2-x}$Sr$_{x}$RuO$_4$ \cite{Fang}, and leads to long-range AFM ordering accompanied by a metal-insulator transtion in Region IV. When \emph{x}$>$0.16, the structure of Sr$_3$(Ru$_{1-x}$Mn$_x$)$_2$O$_7$ is no longer distorted which gives rise to competitive AFM and FM interactions (see Fig. 2(e)). As a result, the system can no longer form long-range magnetic ordering (Region V). On the other hand, the electrical transport is dominated by impurity scattering, leading to the increase of $T_{\rm MIT}$ with \emph{x}. Given the fact that the spin varies from \emph{S}$\sim$1 for \emph{x}=0 to \emph{S}$\sim$3/2 for \emph{x}$>$0.16 (derived from $p_{\rm eff}$ shown in Fig. 2(f)), it is likely that the oxidation state of Mn in Sr$_3$(Ru$_{1-x}$Mn$_x$)$_2$O$_7$ is 4+, independent of \emph{x}. In principle, \emph{S}=3/2 could also result from Ru$^{5+}$, but there seems to be lack of support according to XAS and XPS results \cite{Hossain, Guo}.

In summary, we have investigated the structural and physical properties of Mn-doped Sr$_3$Ru$_2$O$_7$ and constructed a rich phase diagram for 0$\leq$\emph{x}$\leq$0.7. Two characteristic temperatures ($T_{\rm MIT}$ and $T_{\rm M}$) are required to accurately describe the change of the physical properties. $T_{\rm MIT}$ shows a monotonic change, while $T_{\rm M}$ reveals a non-monotonic dependence with \emph{x}. Three distinct regions are identified below $T_{\rm M}$,. This work illustrates the coupling between structure and physical properties which can be tuned by chemical doping.

Work at LSU was partially supported by US National Science Foundation with Grant Nos. DMR-1002622 (B. H., E. W. P., and R. J.) and DMR-1063735 (J. Y. C.). The work at the High Flux Isotope Reactor, Oak Ridge National Laboratory, was sponsored by the Scientific User Facilities Division, Office of Basic Energy Sciences, U.S. Department of Energy.

\end{document}